\documentclass[epjCONF,columns,a4paper]{svjour} 
\usepackage{graphics}
\usepackage[varg]{txfonts} 
\usepackage[latin1]{inputenc}
\usepackage{xspace}
\newcommand{\MET}{\ensuremath{E_{\mathrm{T}}^{\mathrm{miss}}}\xspace}

\newcommand{\mT}{\ensuremath{m_{\mathrm{T}}}\xspace}
\newcommand{\mH}{\ensuremath{m_{\mathrm{H}}}\xspace}
\newcommand{\ifb}{\ensuremath{{\mathrm{fb^{-1}}}}\xspace}
\newcommand{\GeVc}{\ensuremath{{\mathrm{GeV/c}}}\xspace}
\newcommand{\GeVcc}{\ensuremath{{\mathrm{GeV/c^{2}}}}\xspace}
\session-title{The 2011 Hadron Collider Physics symposium (HCP-2011),%
Paris, France, November 14-18 2011}
\begin{document}
\title{Standard Model Higgs Combination from CMS %
  with up to 1.7~${\mathrm{fb^{-1}}}$ of data%
}
\author{Micha\l{} Bluj
\thanks{On leave from NCBJ, Warsaw, Poland}\fnmsep %
\thanks{\email{michal.bluj@cern.ch}} %
for the CMS Collaboration}%
\institute{CNRS-IN2P3/LLR-\'Ecole Polytechnique, Palaiseau, France}
\abstract{%
The combination is presented of searches for a standard model (SM) Higgs boson in 
eight decay modes: $\mathrm{H\!\to\!\gamma\gamma}$, $\mathrm{H\!\to\!\tau\tau}$, 
$\mathrm{H\!\to\!bb}$, $\mathrm{H\!\to\!WW^*\!\to\!2\ell 2\nu}$, 
$\mathrm{H\!\to\!ZZ^*\!\to\!4\ell}$, $\mathrm{H\!\to\!ZZ\!\to\!2\ell 2\tau}$, 
$\mathrm{H\!\to\!ZZ\!\to\!2\ell 2\nu}$, and $\mathrm{H\!\to\!ZZ\!\to\!2\ell 2q}$. 
The searches were performed by the CMS Collaboration using 1.1--1.7~\ifb of 
integrated luminosity, depending on the analysis. No excess compatible with a 
SM Higgs signal has been observed; the largest excursion of the observed data 
from the expected background has a probability of 0.4 after taking into account
 the look-elsewhere effect. The SM Higgs boson is excluded at 95\% C.L. in 
three mass ranges 145--216, 226--288, and 310--400~\GeVcc, while the expected 
exclusion range is 130--440~\GeVcc.%
} 
\maketitle
\section{Introduction}
\label{sec:intro}
Understanding of the mechanism of electroweak symmetry breaking is one of the
main goals of the CMS physics program. In the standard model (SM) the 
electroweak symmetry breaking is described by so-called Higgs mechanism which 
leads to prediction of one scalar particle - a Higgs boson~(H)~\cite{Englert:1964et,Higgs:1964ia,Higgs:1964pj,Guralnik:1964eu,Higgs:1966ev,Kibble:1967sv}. 
So far, experimental searches for this particle have given negative results 
allowing to exclude its mass below $\mathrm{m_H<}$~114.4~\GeVcc (LEP experiments~\cite{Barate:2003sz}) 
and for $\mathrm{m_H\in}$~100--108~\GeVcc and $\mathrm{m_H\in}$~156--177~\GeVcc 
(Tevatron experiments~\cite{CDFandD0:2011aa}). Indirect searches using fits to 
precise measurements predict relatively light Higgs boson with $\mathrm{m_H<}$~158~\GeVcc~\cite{:2010vi}.

In this report combination is presented of searches for the Higgs boson in 
eight decay modes: $\mathrm{H\!\to\! \gamma\gamma}$, $\mathrm{H\!\to\!\tau\tau}$,
$\mathrm{H\!\to\! bb}$, $\mathrm{H\!\to\! WW^*\!\to\! 2\ell 2\nu}$, 
$\mathrm{H\to ZZ^*\to 4\ell}$, $\mathrm{H\to ZZ\to 2\ell 2\tau}$, 
$\mathrm{H\!\to\! ZZ\!\to\! 2\ell 2\nu}$, and $\mathrm{H\!\to\! ZZ\!\to\! 2\ell 2q}$. 

The cross sections for Higgs boson production, its decay branching fractions, 
and their uncertainties are taken from the report prepared by LHC Higgs Cross 
Section group~\cite{Dittmaier:2011ti}.

In Section~\ref{sec:channels} an overview of all eight analysis used in the 
combination is provided, then in Section~\ref{sec:stat} a statistical methodology 
used in this work is briefly described, and finally in Section~\ref{sec:res}
the combined result is presented.
\section{Channels used in the combined search}
\label{sec:channels}
The combination presented in this report bases on eight analyses corresponding 
to main decay modes of the Higgs boson as summarised in Table~\ref{tab:channels}.
\begin{table}[h]
\begin{center}
  \caption{Summary information on the analyses included in the combination. 
  }
  \label{tab:channels}
\scriptsize
\begin{tabular}{ccccc}
\hline
channel                    & mass range & lumi.    & no. of    & type of  \\
                           & [\GeVcc]   & [\ifb]   & sub-chan. & analysis \\
\hline
$H\!\to\!\gamma\gamma$         & 110--150   & 1.7      & 8         & mass shape (unbinned) \\
$H\!\to\!\tau\tau$             & 110--140   & 1.1      & 6         & mass shape (binned)   \\
$H\!\to\! bb$                  & 110--135   & 1.1      & 5         & cut \& count          \\ 
$H\!\to\! WW^*\!\to\! 2\ell 2\nu$  & 110--600   & 1.5      & 5         & cut \& count          \\
$H\!\to\! ZZ^*\!\to\! 4\ell$       & 110--600   & 1.7      & 3         & mass shape (unbinned) \\
$H\!\to\! ZZ\!\to\! 2\ell 2\tau$   & 180--600   & 1.1      & 8         & mass shape (unbinned) \\
$H\!\to\! ZZ\!\to\! 2\ell 2\nu$    & 250--600   & 1.6      & 2         & cut \& count          \\
$H\!\to\! ZZ\!\to\! 2\ell 2q$      & 226--600   & 1.6      & 6         & mass shape (unbinned) \\
\hline
\end{tabular}
\end{center}
\end{table}
In the following subsections a brief description is provided of each individual 
analysis.
\subsection{$\mathrm{H\to \gamma\gamma}$ channel~\cite{CMS-PAS-11-021}} 
\label{sec:gg}%
The $\mathrm{H\!\to\! \gamma\gamma}$ analysis is a search for a narrow peak in the 
di-photon mass $\mathrm{m_{\gamma\gamma}}$ distribution on top of a large falling 
background. It is the most sensitive channel at low masses despite a small 
branching fraction~($\mathrm{Br(H\!\to\!\gamma\gamma)\!\sim\!1\!-\!2\times 10^{-3}}$).

All events are divided into eight categories based on whether the
transverse momentum of the di-photon system \linebreak 
$\mathrm{p_{T}^{\gamma\gamma}}>$~40~\GeVc, 
whether both photons are in the central part of the CMS detector, and whether 
both photons are unconverted (have compact electromagnetic showers). The 
categorisation is motivated by different resolution in each category~(1-3\%). 

The background under the expected signal peak is derived from sidebands without
use of Monte Carlo simulation.
\subsection{$\mathrm{H\to\tau\tau}$ channel~\cite{CMS-PAS-11-009}} 
\label{sec:tt}%
In this analysis a broad excess in the visible di-tau mass 
$\mathrm{m_{\tau\tau}^{vis}}$ distribution is looked for (resolution of 
$\mathrm{m_{\tau\tau}^{vis}}\sim$~20\%). 

Three di-tau final states are used: $\mathrm{e\mu}$, $\mathrm{e\tau_{h}}$, 
$\mathrm{\mu\tau_{h}}$ ($\mathrm{\tau_{h}}$~stands for a $\mathrm{\tau}$ 
decaying hadronically). Each of of these three categories is further divided 
into two mutually exclusive sub-categories: events with the Vector Boson Fusion
 (VBF) signature (two jets separated in pseudorapidity with no additional jets 
in between), and events with less than two jets or with exactly two jets that 
fail VBF requirements. 

The main irreducible background is {$\mathrm{Z\!\to\!\tau\tau}$
with normalisation taken from the $\mathrm{Z\!\to\!\ell\ell}$ cross section measurement and
shape of $\mathrm{m_{\tau\tau}^{vis}}$ modelled using simulated events. 
The reducible backgrounds (W+jets, QCD, $\mathrm{t\bar{t}}$, 
$\mathrm{Z\!\to\!\ell\ell}$) are evaluated basing on data.
\subsection{$\mathrm{H\to bb}$ channel~\cite{CMS-PAS-11-011}}
\label{sec:bb}%
The $\mathrm{H\to bb}$ search exploits the Higgs boson production in association 
with W or Z bosons~(V). The following W and Z boson decay modes are considered:
 $\mathrm{W\to\ell\nu}$ and $\mathrm{Z\to\ell\ell,\nu\nu}$ ($\mathrm{\ell=e,\mu}$).
It is required that the system of two b-tagged jets (a Higgs boson decay candidate)
is boosted in the transverse plane, which reduces background and improves the 
di-jet mass resolution. 

The result of the analysis bases on event count in signal regions defined by the
output of a multivariate analysis classifier (MVA). The classifier was trained 
for a number of Higgs boson masses. 

The rates of the main backgrounds V+jets, $\mathrm{Vb\bar{b}}$, and 
$\mathrm{t\bar{t}}$ are estimated from control samples and then applied to 
simulation. The WZ and ZZ backgrounds with $\mathrm{Z\!\to\! b\bar{b}}$, and the 
single-top background are modelled with Monte Carlo simulation.
\subsection{$\mathrm{H\to WW^*\to 2\ell 2\nu}$ channel~\cite{CMS-PAS-11-014}}
\label{sec:WW}%
The signature of the $\mathrm{H\!\to\! WW^*\!\to\! 2\ell 2\nu}$ signal is the
presence of exactly two opposite~sign, isolated leptons and significant \MET. 
There is no signal mass pick due to escaping energy due to presence of two 
neutrinos from W~decays. The search is based on event counting. 

Events are split into three categories basing on a jet multiplicity in the event 
(0,~1 or 2~jets) with different signal-to-background ratios. For the 0-jet 
category the main background is the electroweak WW production; for the 1-jet category 
the WW and {$\mathrm{t\bar{t}}$ processes. Both the 0- and 1-jet categories are further split 
into same-flavour and opposite-flavour di-lepton sub-channels, since different
contribution of the Drell-Yan background. The 2-jet category is optimised to 
take advantage of the VBF production signature (jets separation in pseudorapidity). 
The main background for this category is {$\mathrm{t\bar{t}}$.

To separate the $\mathrm{H\!\to\! WW}$ signal from the electroweak WW background 
the scalar nature of the Higgs boson is explored.

Contributions from all main backgrounds are estimated basing on data.
\subsection{$\mathrm{H\to ZZ^*\to 4\ell}$ channel~\cite{CMS-PAS-11-015}}
\label{sec:4l}
The $\mathrm{H\!\to\! ZZ^*\!\to\! 4\ell}$ analysis is a search for a four-lepton mass 
peak over the continuum background. Three final states 4e, 4$\mathrm{\mu}$, 
2e2$\mathrm{\mu}$ are considered separately, as each of them has a different 
resolution of $\mathrm{m_{4\ell}}$ and a different composition of background with 
jets faking leptons. 

The dominant irreducible background is the electroweak ZZ production. Its contribution
is modelled using simulation and normalised using the measured yield of 
$\mathrm{Z\to\ell\ell}$ events scaled by the ratio of ZZ and Z cross sections.
The reducible backgrounds with jets faking leptons (Z+jets, $\mathrm{Zb\bar{b}}$,
 $\mathrm{t\bar{t}}$) are evaluated from data using control regions. Their 
contribution was found to be small.
\subsection{$\mathrm{H\to ZZ\to 2\ell 2\tau}$ channel~\cite{CMS-PAS-11-013}}
\label{sec:2l2t}
In the $\mathrm{H\!\to\! ZZ\!\to\! 2\ell 2\tau}$ search, the presence is required of one 
di-lepton pair (ee or $\mathrm{\mu\mu}$) forming an on-shell Z boson. Then 
a second Z boson is required to decay to $\mathrm{\tau}$-pair, with one of the 
following four decay modes: $\mathrm{e\mu}$, $\mathrm{e\tau_{h}}$, 
$\mathrm{\mu\tau_{h}}$, $\mathrm{\tau_{h}\tau_{h}}$ ($\mathrm{\tau_{h}}$~stands 
for a $\mathrm{\tau}$ decaying hadronically). It makes eight exclusive 
sub-channels. In the analysis, the mass of two leptons and visible products of 
two tau decays (without accounting for missing neutrinos) is a final observable.

The dominant background is the electroweak ZZ production which is taken from 
simulation and normalised using the measured yield of $\mathrm{Z\!\to\!\ell\ell}$ events 
scaled by the ratio of ZZ and Z cross sections. The sub-leading backgrounds with 
jets faking tau come from Z+jets (including ZW) and $\mathrm{t\bar{t}}$, are 
evaluated from data using fake-rate method.
\subsection{$\mathrm{H\to ZZ\to 2\ell 2\nu}$ channel~\cite{CMS-PAS-11-016}}
\label{sec:2l2nu}
In the $\mathrm{H\!\to\! ZZ\!\to\! 2\ell 2\nu}$ analysis, events with one di-lepton pair 
(ee or $\mathrm{\mu\mu}$) consistent with an on-shell Z boson, and significant 
\MET are selected. Then the transverse mass \mT from the di-lepton pair momenta
 and \MET is constructed assuming that \MET arises from the $\mathrm{Z\!\to\!\nu\nu}$ 
decay\footnote{$\mathrm{m_T^2=\left(\sqrt{p_{T,Z}^2+M_Z^2}+\sqrt{(E_T^{miss})^2+M_Z^2}\right)^2-\left(\overrightarrow{p}_{T,Z}+\overrightarrow{E}_T^{miss}\right)^2}$}. Finally, events are counted in a \mH dependent window in the \mT distribution. 

The main ZZ and WZ backgrounds are taken from simulation, while all other 
backgrounds are evaluated from control samples.
\subsection{$\mathrm{H\to ZZ\to 2\ell 2q}$ channel~\cite{CMS-PAS-11-017}}
\label{sec:2l2q}
The $\mathrm{H\!\to\! ZZ\!\to\! 2\ell 2q}$ analysis is a search for a peak in the mass 
of the di-lepton plus di-jet system~($\mathrm{m_{2\ell 2j}}$).
There are six exclusive final states used in the search with the lepton pair in one
of two possible flavours (ee or $\mathrm{\mu\mu}$) and the jet pair with 0, 1 or 
2 b-tags. Both lepton and jet pairs are required to be consistent with the 
Z boson mass. Background is further suppressed by employing a multivariate 
angular likelihood constructed from the kinematic variables of the two leptons 
and the two jets. 

The background $\mathrm{m_{2\ell 2j}}$ distribution is obtained using control 
regions in data.
\section{Statistical methodology}
\label{sec:stat}
The modified frequentist construction CLs~\cite{Read:2000ru,Junk:1999kv} is used
for calculations of exclusion limits presented in this report. To completely 
define the method, one needs to make a choice of test statistic and of 
treatment of the systematic uncertainties in the construction of the test 
statistic and in generating pseudo-data. Here the LHC Higgs Combination Group 
prescription~\cite{LHC-HCG-Report:2011} is followed.

The likelihood $\mathcal{L}(data\,|\,\mu,\theta )$ used to construct the 
test statistic is defined as follows:
\begin{equation}
\label{eq:LHC-Likelihood}
\mathcal{L}(data|\,\mu,\theta) = \mathrm{Poisson}\left(data|\,\mu\cdot s(\theta)+b(\theta)\right)\cdot p(\tilde{\theta}|\theta)\, ,
\end{equation}
where $\mathrm{Poisson}\left(data|\mu\cdot s(\theta)+b(\theta)\,\right)$ is 
the Poisson probability to observe ``data'', assuming signal and background  
models, $s(\theta)$ and $b(\theta)$, that depend on some nuisance parameters 
$\theta$. The free parameter $\mathrm{\mu}$ is a common signal strength 
modifier affecting signal event yields in all production modes ($\mathrm{\sigma/\sigma_{SM}}$). Nuisance parameters $\theta$ correspond with independent sources 
of systematic uncertainties. The probability of ``measuring'' $\mathrm{\tilde{\theta}}$ 
which is best known estimate of true value $\mathrm{\theta}$ is 
$\mathrm{p(\tilde{\theta}|\theta)}$ and describes the scale of the systematic 
uncertainty.

Then, the test statistics is definied as:
\begin{equation}
\label{eq:qmu}
q_{\mu}\,\,=\,\, 
-2\ln\frac{\mathcal{L}(\mathrm{data}|\mu,\hat{\theta}_{\mu})}{\mathcal{L}(\mathrm{data}|\hat{\mu},\hat{\theta})}, \hspace{0.5cm} \textrm{where} 0\le\hat{\mu}\le\mu,
\end{equation}
where ``data'' stands either for the real observation or pseudo-data. 
Both the denominator and numerator are maximized. In the numerator, $\mathrm{\mu}$ 
is fixed and only the nuisance parameters $\mathrm{\theta}$ can float. Their 
values at which $\mathcal{L}$ reaches the maximum are denoted as 
$\mathrm{\hat\theta_{\mu}}$. In the denominator, both $\mathrm{\mu}$ and 
$\mathrm{\theta}$ are allowed to float in the fit, and $\mathrm{\hat\mu}$ and 
$\mathrm{\hat\theta}$ are parameters at which $\mathcal{L}$ reaches its global 
maximum. The lower constraint on $\hat\mu$ ($\mathrm{0\leq\hat\mu}$) is imposed 
as the signal rate cannot be negative; the upper constraint ($\mathrm{\hat\mu\leq\mu}$) forces the limit to be one-sided. The value of the test statistic for 
the real observation is denoted as $\mathrm{q_{obs}}$.

In the next step, the values of the nuisance parameters $\mathrm{\hat\theta^{obs}_{0}}$ 
and $\mathrm{\hat\theta^{obs}_{\mu}}$ best describing the observed data (maximizing 
$\mathcal{L}$) are obtained for the background-only and signal+background hypotheses, 
respectively. Using these best-fit values of the nuisance parameters, toy Monte 
Carlo pseudo-data is generated to construct the test statistic sampling distributions 
for the both signal+background hypothesis (with signal strength $\mu$) and for 
the background-only hypothesis ($\mu$=0). The ``measurements'' $\mathrm{\tilde{\theta}}$ 
are also randomized in each pseudo-data.

With signal+background and background-only sampling distributions for the test 
statistic $\mathrm{q_\mu}$ one can find the probability to obtain a test 
statistic value as high as, or higher than, the one observed in data, under 
the signal+background hypothesis, and obtain CLs($\mu$) from the ratio
\begin{equation}
  CL_s(\mu)=\frac{P\left(q_{\mu}\geq q_{\mu}^{obs}|\,\mu\cdot s(\hat\theta^{obs}_{\mu})+ b(\hat\theta^{obs}_{\mu})\right)}
  {P\left(q_{\mu}\geq q_{\mu}^{obs}|\,b(\hat\theta^{obs}_0)\right)}.
\end{equation}
The $\mathrm{CLs\leq\alpha}$ for a given $\mu$ means that the signal with 
strenth~$\mu$ is excluded at the ($\mathrm{1-\alpha}$) confidence level (C.L.). 
To quote the 95\% C.L. upper limit on $\mu$, we adjust $\mu$ until we reach 
$\mathrm{CLs=1-0.95}$.

To quantify an excess of events the alternative test statistic $\mathrm{q_0}$ 
is used:
\begin{equation}
\label{eq:qmu0}
q_{0}=
-2\ln\frac{\mathcal{L}(\mathrm{data}\,|\,0,\hat{\theta}_0)}
          {\mathcal{L}(\mathrm{data}\,|\,\hat{\mu},\hat{\theta})} \hspace{0.5cm} 
\textrm{and }\hat{\mu}\geq 0.
\end{equation}
This test statistic allows to evaluate significances (Z) and p-values ($\mathrm{p_0}$) from the asymptotic formula:
\begin{equation}
\label{eq:Z}
Z = \sqrt{ q_0^{\mathrm{obs}} },
\end{equation}
\begin{equation}
\label{eq:p-estimate}
p_0  = P(q_0 \geq q_0^{\mathrm{obs}}) =\!\int_Z^{\infty} \frac {e^{-x^2/2} } {\sqrt{2\pi}}\, dx =  \frac{1}{2} \left[ 1 - \mathrm{erf} \left( Z/\sqrt{2} \right) \right],
\end{equation}
where $\mathrm{q^{obs}_0}$ is the observed test statistic for $\mu$~=~0.
\section{Result of combined search}
\label{sec:res}
The combined result of the search for the SM boson is presented in Figure~\ref{fig:Comb} 
which shows the combined 95\% C.L. upper limits on the signal strength modifier 
($\mathrm{\sigma/\sigma_{SM}}$) as a function of the Higgs boson mass. The 
dashed line stands for the median expected results for the background-only 
hypothesis, while the green (yellow) bands indicate the range of 68\% (95\%) 
deviation from the median. The observed limit is indicated by the solid line 
with points. The SM Higgs boson is excluded at 95\%~C.L. in three mass ranges 
145--216, 226--288, and 310--400~\GeVcc, while the expected exclusion range is 
130--440~\GeVcc.

\begin{figure}[h]
\resizebox{1.00\columnwidth}{!}{%
  \includegraphics{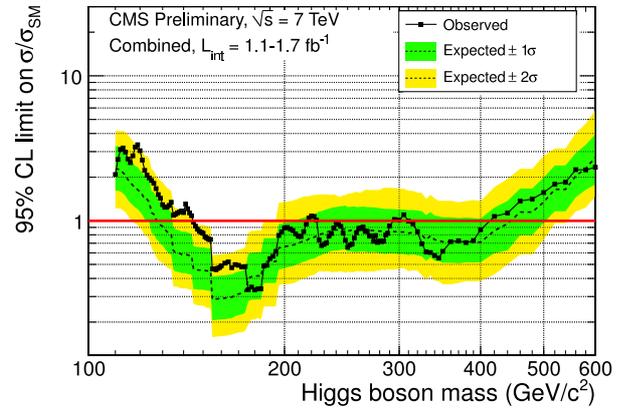} }
\caption{The 95\% C.L. upper limit on the signal cross section 
  $\mathrm{\sigma/\sigma_{SM}}$ for the SM Higgs boson hypothesis as a function 
  of the SM Higgs boson mass. The observed values are shown by the solid line. 
  The dashed black line indicates the median expected results for the background-only
  hypothesis, while the green (yellow) bands indicate the range of 68\% (95\%) 
  deviation from the median.}
\label{fig:Comb}
\end{figure}
A contribution of each individual Higgs decay mode to the combined result is 
illustrated in Figure~\ref{fig:CombSplit}. For high masses ($\mathrm{m_H>}$200~\GeVcc) 
the exclusion is driven by the $\mathrm{H\!\to\! ZZ}$ decay channels, for 
an intermediate range 130--200~\GeVcc the exclusion is dominated 
by the $\mathrm{H\!\to\! WW^*\!\to\! 2\ell 2\nu}$ mode, while for low masses 
($\mathrm{m_H<}$130~\GeVcc) $\mathrm{H\!\to\!\gamma\gamma}$ is the most significant 
contributor.

\begin{figure}[h]
\resizebox{1.00\columnwidth}{!}{%
  \includegraphics{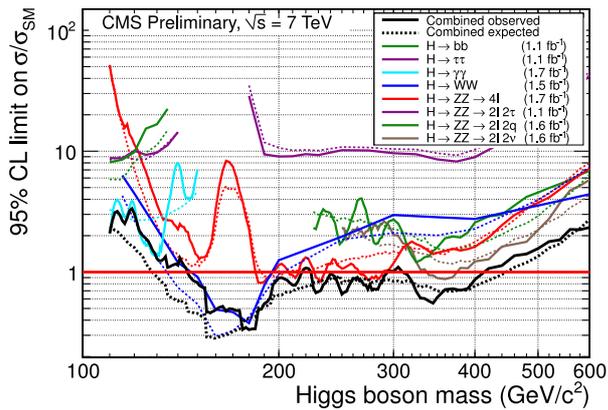} }
\caption{The observed 95\% C.L. upper limit on the signal cross section 
  $\mathrm{\sigma/\sigma_{SM}}$ for the SM Higgs boson hypothesis as a function 
  of the SM Higgs boson mass for eight decay modes and their combination.}
\label{fig:CombSplit}
\end{figure}
The differences between the observed and expected limits are consistent
with statistical fluctuations, as the observed limits lie within the 68\% and 
95\% bands. For the low Higgs boson mass range, we observe an excess of events which
leads to weaker observed limit than expected in the absence of the SM Higgs boson.
The observed local p-value~$\mathrm{p_0}$ which quantifies the consistency of 
the observed excesses with the background-only hypothesis, is shown in 
Figure~\ref{fig:pO} (top panel) for the combined search, and split into 
individual decay modes in Figure~\ref{fig:p0Split}. A broad offset for low 
masses, of about 1$\mathrm{\sigma}$, corresponds to the excesses seen in the 
$\mathrm{H\!\to\! WW^*\!\to\! 2\ell 2\nu}$ channel characterised by poor mass 
resolution. The excesses observed in the $\mathrm{H\!\to\!ZZ\!\to\!4\ell}$ and 
the $\mathrm{H\!\to\!\gamma\gamma}$ channels result two narrow minima of the p-value. 
The minimal p-value is $\mathrm{p_{min}\sim\!0.01}$, but after accounting for 
the look-elsewhere effect which is important for this study, it is reduced to a 
global probability $\mathrm{p_{global}\!\sim\!0.4}$. The look-elsewhere effect was
 esimeted for the whole explored mass range 110--600~\GeVcc. 

The best-fit value of $\mathrm{\sigma/\sigma_{SM}}$ is also presented in 
Figure~\ref{fig:pO} (bottom panel). The best-fit value is a factor by which the 
SM Higgs boson cross section has to be rescaled to best describe observed data.
In the mass range between 115 and 125~\GeVcc the best-fit value agrees within
uncertainty with 1 ($\mathrm{\sigma\!=\!\sigma_{SM}}$), but as discussed above, 
with the analysed amount of data the excess is not significant. More data will 
increase the statistical accuracy of the search thus improve its sensitivity. 

\begin{figure}[h]
\resizebox{1.00\columnwidth}{!}{%
  \includegraphics{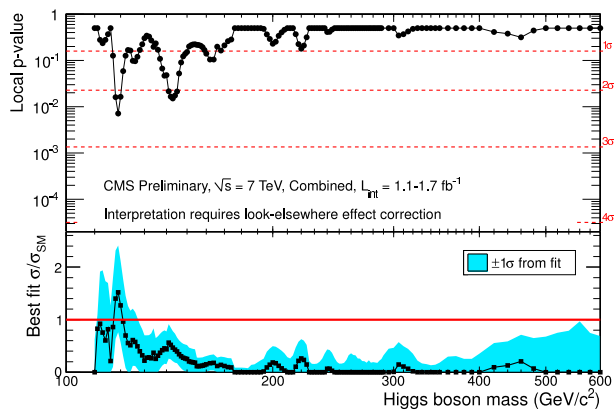} }
\caption{The observed local p-value $\mathrm{p_0}$ (top panel) and 
  $\mathrm{\sigma/\sigma_{SM}}$ of the best-fit (bottom panel) as a function of
  the SM Higgs boson mass. The maximal observed excess (minimal local p-value) 
  corresponds with a global probability for background-only hypothesis equal to 
  0.4 after accounting for the look-elsewhere effect for $\mathrm{m_H}$=110--600~\GeVcc.}
\label{fig:pO}
\end{figure}
\begin{figure}[h]
\resizebox{1.00\columnwidth}{!}{%
  \includegraphics{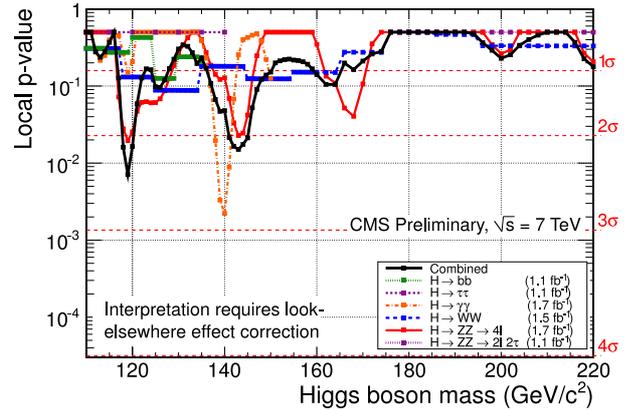} }
\caption{The observed local p-value as a function of the SM Higgs boson mass 
  for eight decay modes and their combination.}
\label{fig:p0Split}
\end{figure}
\section{Conclusions}
\label{sec:concl}
The combined search for the standard model Higgs boson performed by the CMS 
Collaboration with up to 1.7~\ifb of data was presented. The expected exclusion 
mass range is 130--440~\GeVcc. The observed data allowed to exclude the SM 
Higgs boson at 95\% C.L. in three mass ranges 145--216, 226--288, and 
310--400~\GeVcc. The largest excursion of the observed data from the expected 
background has a probability of 0.4 after taking into account the 
look-elsewhere effect for the whole explored mass range (110--600~\GeVcc).
%

%
\end{document}